\documentclass[12pt]{article}
\usepackage{graphicx}

\def\ignore#1{{}}

\let\oldtheequation=\theequation
\def\doteqs#1{\setcounter{equation}{0}            
\def\theequation{{#1}.\oldtheequation}}
\newcounter{sxn}
\def\sx#1{\addtocounter{sxn}{1} \vskip 1.cm  \goodbreak
\noindent{\large\bf\leftline{\thesxn.~~#1}} \nobreak \vskip -.5cm}
\def\sxn#1{\sx{#1} \doteqs{\thesxn}}

\newcounter{axn}

\date{}

\newdimen\mybaselineskip
\newcommand{\beeq}{\begin{equation}}
\newcommand{\eneq}{\end{equation}}
\newcommand{\beqn}{\begin{eqnarray}}
\newcommand{\eeqn}{\end{eqnarray}}

\def\la{\raise.16ex\hbox{$\langle$}\lower.16ex\hbox{}  }
\def\ra{\, \raise.16ex\hbox{$\rangle$}\lower.16ex\hbox{} }

\def\psibar{ \psi \kern-.65em\raise.6em\hbox{$-$} \lower.6em\hbox{} }
\def\psibarb{ \psi \kern-.65em\raise.6em\hbox{$-$}  }

\begin{document}

\thispagestyle{empty}

\baselineskip=12pt



\vspace*{3.cm}

\begin{center}  
{\LARGE \bf  The Mystery of the Asymptotic Quasinormal Modes of Gauss-Bonnet Black Holes}
\end{center}

\baselineskip=14pt

\vspace{3cm}
\begin{center}
{\bf  Ramin G. Daghigh $\sharp$, Gabor Kunstatter $\dagger$, and Jonathan Ziprick $\dagger$}
\end{center}

\centerline{\small \it $\sharp$ Natural Sciences Department, Metropolitan State University, Saint Paul, Minnesota, USA 55106}
\vskip 0 cm
\centerline{} 

\centerline{\small \it $\dagger$ Physics Department, University of Winnipeg, Winnipeg, Manitoba, Canada R3B 2E9}
\centerline{\small \it and}
\centerline{\small \it Winnipeg Institute for Theoretical Physics, Winnipeg, Manitoba}
\vskip 1 cm
\centerline{} 

\vspace{1cm}
\begin{abstract}
We analyze the quasinormal modes of $D$-dimensional Schwarzschild black holes with the Gauss-Bonnet correction in the large damping limit and show that standard analytic techniques cannot be applied in a straightforward manner to the case of infinite damping. However, by using a combination of analytic and numeric techniques we are able to calculate the quasinormal mode frequencies in a range where the damping is large but finite.  We show that for this damping region the famous $\ln(3)$ appears in the real part of the quasinormal mode frequency.  In our calculations, the Gauss-Bonnet coupling, $\alpha$, is taken to be much smaller than the parameter $\mu$, which is related to the black hole mass. 
\baselineskip=20pt plus 1pt minus 1pt
\end{abstract}

\newpage

\sxn{Introduction}

\vskip 1cm
Although the highly damped quasinormal modes (QNMs) of black holes are not directly observable, these modes have been the subject of intense investigation since the observation \cite{Hod, Dreyer} that they may in principle provide information about the semi-classical quantum spectrum of black holes. This observation was motivated in large part by the special value of the QNM frequencies of Schwarzschild black holes in the high damping regime, namely:
\beeq
\omega_R \mathop{\longrightarrow}_{|\omega_I|\to\infty}\ln(3)T_{bh}/\hbar ~,
\eneq
where $\omega_I$ is the damping and $T_{bh}$ is the black hole temperature.  We refer the reader to the literature for details \cite{Natario}. The all important $\ln(3)$ in the coefficient is present for a large class of single horizon black holes \cite{Motl2,Gabor3,Das1,Ramin1} but not for multi-horizon black holes. Since the arguments that relate the QNM frequencies to black hole spectra are strictly speaking only valid as given in the single horizon case, it is of great interest to ask whether the $\ln(3)$ persists for all possible single horizon black holes of physical interest. This provides the motivation for the following analysis of Schwarzschild black holes with Gauss-Bonnet (G-B) corrections.  In the rest of this paper, we will refer to these black holes as G-B black holes.

G-B black holes \cite{GB1, GB2, GB3} are black holes in spacetime dimensions greater than four which appear when one includes, in the Einstein-Hilbert action, the leading order higher curvature terms arising in the low energy limit of string theories \cite{Zwiebach}.  These black holes have recently attracted considerable attention due to the possibility of their production at the Large Hadron Collider if the Planck scale is of the order of one TeV \cite{GB-brane} as occurs in extra dimensional brane-world scenarios.  

Reference \cite{GB-qnm1} for the first time discusses the scalar QNMs in the background of G-B black holes and evaluates these modes for a few values of the G-B coupling in five and six dimensions.  The scalar QNMs in the background of G-B and charged G-B black holes have been studied in \cite{GB-qnm2} and this work has been extended to include G-B-de Sitter and G-B-Anti-de Sitter backgrounds in \cite{GB-qnm3}.  The QNMs for tensor perturbations of G-B black holes in $5$, $7$, and $8$ dimensions were calculated in \cite{GB-qnm4} using a third order WKB formalism.

More recently, there has been an attempt by Chakrabarti and Gupta \cite{GB-asymp-qnm} to obtain an analytic expression for the highly damped QNM frequencies of G-B black holes.  These authors calculated the highly damped frequencies of tensorial perturbations in the limit where ${\alpha \over r} \ll 1$ in which $\alpha$ is the G-B coupling constant and $r$ is the radial coordinate.  This approximation, however, is not valid in the highly damped limit where we need to extrapolate the WKB solutions to a region where $r\sim 0$. The flaw in the results of \cite{GB-asymp-qnm} has also been pointed out by Moura and Schiappa  \cite{GB-asymp-qnm1}.  

The purpose of the present paper is to analyze the QNMs of $D$-dimensional G-B black holes in the large damping limit. We will show that standard analytic techniques cannot be applied in a straightforward manner to the case of infinite damping. However, by using a combination of analytic and numeric techniques we calculate the QNM frequencies in a range where the damping is large but finite. 

In Sec. 2, we introduce briefly the G-B black hole and describe the general formalism for calculating QNMs. The specific equations describing the vector perturbations of G-B black holes are presented in Sec. 3, where we argue that the standard analytic techniques cannot extract the infinitely damped modes. In Sec. 4 we calculate the corresponding QNMs in the ``intermediate'' damping region. The last section closes with conclusions and prospects for further work.

\sxn{Basic Formalism}

The Einstein-Hilbert action in the presence of the G-B term has the form 
\beeq
I ={1\over 16\pi G_D}\left[\int d^Dx\sqrt{-g}R+{\alpha \over(D-3)(D-4)}\int d^Dx\sqrt{-g}\left(R_{abcd}R^{abcd}-4R_{cd}R^{cd}+R^2\right)\right]~.
\label{GB-action}
\eneq
where $D\geq5$
and $\alpha$ is the G-B coupling constant which is positive valued and is related to the Regge slope parameter or string scale.  The metric for the spherically symmetric, asymptotically flat black hole solution of the action (\ref{GB-action}) is \cite{Deser}
\beeq
ds^2=-f(r)dt^2+{dr^2 \over f(r)}+r^2d\Omega^2_{D-2}~,
\label{GB-metric}
\eneq
where 
\beeq
f(r)=1+{r^2\over 2\alpha}-{r^2\over 2\alpha}\sqrt{1+{8\alpha\mu \over r^{D-1}}}~.
\label{GB-f}
\eneq
The ADM mass, $M$, of the black hole is related to the parameter $\mu$ by
\beeq
M = {(D-2)A_{D-2} \over 8 \pi G_D}\mu~,
\eneq
where $A_n$ is the area of a unit $n$-sphere,
\beeq
A_n={2\pi^{n+1 \over 2} \over \Gamma\left({n+1 \over 2}\right) }~.
\eneq

The QNM perturbations of black holes are governed generically by a Schr$\ddot{\mbox o}$dinger wave-like equation of the form
\beeq
{d^2\psi \over dz^2}+\left[ \omega^2-V(r) \right]\psi =0 ~,
\label{Schrodinger}
\eneq
for perturbations that depend on time as $e^{-i\omega t}$. The potential $V$ depends in general on the black hole solution as well as the nature of the perturbation. It is useful to define the tortoise coordinate $z$ by 
\beeq
dz ={dr \over f(r) }~.
\label{tortoise}
\eneq
For the cases of interest here, the effective potential vanishes at both the event horizon ($z\rightarrow -\infty$) and spatial infinity ($z\rightarrow \infty$), so that the asymptotic behavior of the QNM solutions is chosen to be
\beeq
\psi(z) \approx \left\{ \begin{array}{ll}
                   e^{-i\omega z}  & \mbox{as $z \rightarrow -\infty$ $(r\rightarrow r_h)$}~,\\
                   e^{+i\omega z}  & \mbox{as $z\rightarrow \infty$ $(r\rightarrow \infty)$~,}
                   \end{array}
           \right.        
\label{asymptotic}
\eneq 
which represents an outgoing wave at infinity and an ingoing wave at the horizon.  Since the tortoise coordinate is multi-valued, it is more convenient to work in the complex $r$-plane.  After rescaling the wavefunction $\psi=\Psi/\sqrt{f}$ we obtain
\beeq
\frac{d^2\Psi}{dr^2}+R(r)\Psi=0 ~,
\label{Schrodinger-r}
\eneq 
where
\beeq
R(r)= {\omega^2\over f^2(r)}-U(r)~,
\label{Rr}
\eneq
with
\beeq
U(r)={V(r)\over f^2}+{1\over 2}\frac{f''}{f}-\frac{1}{4}\left(\frac{f'}{f}\right)^2 ~.
\label{Ur}
\eneq
Here prime denotes differentiation with respect to $r$.

\sxn{Vector perturbations and the infinite damping limit}

It was shown by Dotti and Gleiser in \cite{Dotti1}, \cite{Dotti2}, and \cite{Dotti3} that various classes of static, spherically symmetric black hole metric (tensor, vector, and scalar) perturbations in the presence of G-B term in a spacetime with dimension $D>4$ are governed generically by a Schr$\ddot{\mbox o}$dinger wave-like equation of the form (\ref{Schrodinger}).
In this paper we focus on vector perturbations because, tensor and scalar perturbations cannot be done using the numerical techniques employed in this paper: a crucial term in the effective potential vanishes precisely in the intermediate damping region studied in the next section.  

The effective potential for vector perturbations is given by \cite{Dotti3}
\beeq
V_v(r)= q_v+{1 \over K_v}\left(f^2 {d^2K_v \over dr^2}+f f' {dK_v \over dr} \right) ~,
\label{Vv}
\eneq
with $K_v(r)$ and $q_v(r)$ being given by
\beeq
K_v(r)=\left \{r^{D-2}+{2 \alpha r^2 \over D-3}{d \over dr}\left[r^{D-1}(1-f)\right] \right\}^{-1/2} ~
\label{Kv}
\eneq
and
\beeq
q_v(r)={(k_v^2-D+3)f \over r^2}\left\{ 2 \alpha^2 {D-5\over D-3}(1-f)^2 + 2 \alpha {D-5 \over D-3}r^2(1-f)+r^4 \over [2 \alpha (1-f)+r^2]^2 \right \}~,
\label{qv}
\eneq
where $k_v^2=l(l+n-1)-1$ with $l=2,3,4,\cdots$ .

We now consider the QNMs in the infinite damping limit where
\beeq
|\omega^2|\to [{\rm Im}~\omega]^2 \to \infty~.
\eneq
Since we are using complex analytic techniques, in principle the behavior of $U(r)$ on the entire complex plane may be relevant.
However, in the infinite damping limit case, the $\omega^2/f^2$ term in $R(r)$ will dominate $U(r)$ everywhere, unless one of the terms in $U(r)$ diverges.  This can only happen at the origin, the event horizon, the extra fictitious (i.e. complex) horizons, and in the case of G-B black holes at the poles where $r^{D-1}+8\alpha \mu=0$ and $(D-3)r^{D-1}+4\alpha \mu(D-5)=0$.  Note that the poles at the points where $(D-3)r^{D-1}+4\alpha \mu(D-5)=0$ only exist in spacetime dimensions greater than five.  Since $\omega^2/f^2$ also diverges at the event horizon and extra fictitious horizons, it will dominate there as well in the large damping limit. Thus, only the dominant terms of $U(r)$ near the origin and the poles where $r^{D-1}+8\alpha \mu=0$ and $(D-3)r^{D-1}+4\alpha \mu(D-5)=0$ are relevant in this limit.  Therefore, for example in five spacetime dimensions, $R(r)$ can be approximated on the entire complex plane by
\beeq
R(r)\sim \frac{\omega^2}{f^2} -{35 \over 4r^2}+ {48 \alpha \mu \over (r^4+8 \alpha \mu)^{3/2}}~.
\label{Rr5d}
\eneq
In higher spacetime dimensions this expression becomes more complicated.

In the WKB analysis, the two solutions to Eq. (\ref{Schrodinger-r}) are
approximated by
\beeq
\left\{ \begin{array}{ll}
                   \Psi_1^{(t)}(x)=Q^{-1/2}(x)\exp \left[+i\int_{t}^xQ(x')dx'\right]~,\\
                   \\
                   \Psi_2^{(t)}(x)=Q^{-1/2}(x)\exp \left[-i\int_{t}^xQ(x')dx'\right]~,
                   \end{array}
           \right.        
\label{WKB}
\eneq
where
\beeq
Q^2(r)=R(r)-\frac{1}{4r^2}\sim \frac{\omega^2}{f^2}-\frac{9}{r^2}+ {48 \alpha \mu \over (r^4+8 \alpha \mu)^{3/2}}~
\label{Q^2-5d}
\eneq
is shifted by $1/(4r^2)$ in order to guarantee the correct behavior of the WKB solutions at the origin.  In Eq. (\ref{WKB}), $t$ is a simple zero of $Q^2$.

\begin{figure}[tb]
\begin{center}
\includegraphics[height=9cm]{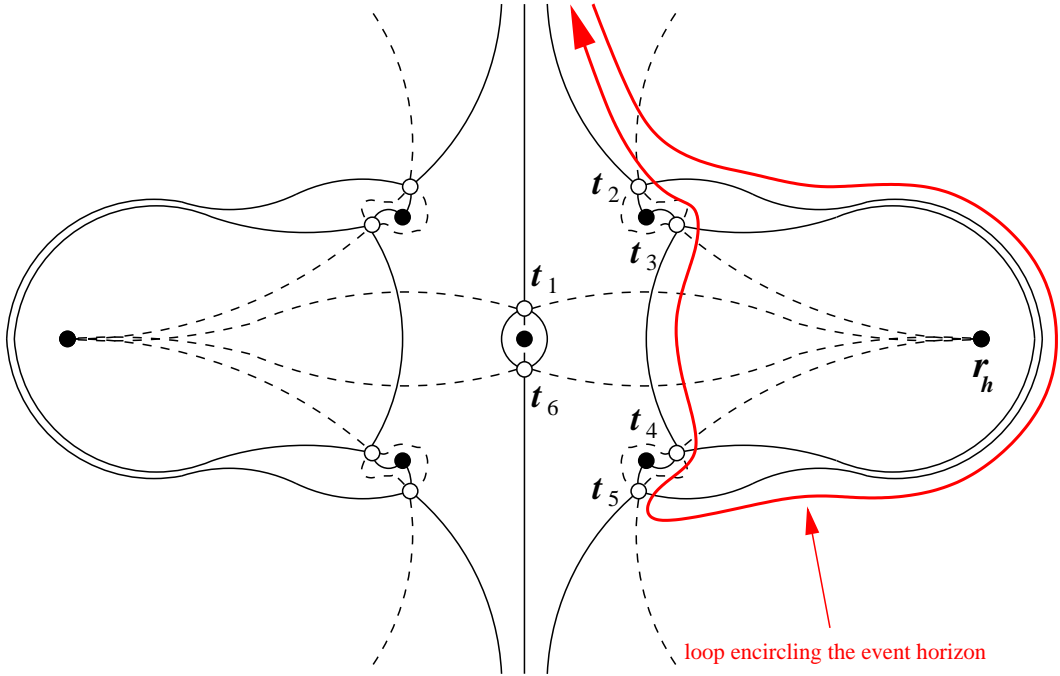}
\end{center}
\caption{A schematic illustration of the Stokes (dashed) and anti-Stokes (solid) lines in the complex $r$-plane for the G-B problem in five spacetime dimensions in the infinite damping limit.  The open circles represent the zeros, while the filled circles are the poles of $Q^2$.}
\label{schematicGB}
\end{figure}

In order to find the QNM frequency, $\omega$, first we need to determine the zeros and poles of the function $Q$ and the subsequent behavior of the Stokes and anti-Stokes lines in the complex $r$-plane.  Stokes lines are the lines on which the WKB phase ($\int Q dr$) is purely imaginary and anti-Stokes lines are the lines on which the WKB phase is purely real.  Due to the appearance of extra poles in G-B corrected Schwarzschild spacetimes, the topology of the Stokes/anti-Stokes lines in the complex plane will look very different from the topology of these lines in the Schwrzschild problem.  A schematic behavior of Stokes/anti-Stokes lines for the G-B problem in the infinite damping limit is shown in Fig. \ref{schematicGB} for five spacetime dimensions. Although this figure was drawn for $\alpha$ small compared to the horizon scale, the topology does not change as $\alpha$ is increased. The extra poles merely move outward in the complex plane as illustrated in Fig. \ref{schematicGB}. 

Once we determine the structure of Stokes/anti-Stokes lines, we can follow the analytic method used in \cite{Andersson} to extract the WKB condition which consequently determines the QNM frequencies in the highly damped limit.  In this method one needs to find a contour that starts and ends on an anti-Stokes line that extends to infinity in the complex plane, encircling only the horizon.  The solution at infinity is fixed due to the boundary condition at infinity and the monodromy of the solution along the entire contour can be calculated using standard (``Stokes phenomena'') rules, provided the contour stays along anti-Stokes lines everywhere except perhaps near zeros of $Q^2$. This monodromy must equal the monodromy close to the horizon, which is in effect fixed by the boundary condition near the horizon. From this equality one gets an algebraic condition that determines the QNM frequency spectrum. One potential loop of this type is shown in Fig. \ref{schematicGB}.  

Unfortunately, it seems impossible to extract a WKB condition on the highly damped QNM frequency given the topology shown in Fig. \ref{schematicGB}.  The reason is that any closed contour beginning and ending at infinity, and circling only the horizon necessarily passes through the extra poles which appear due to the G-B correction.  This happens for example when we move along the anti-Stokes lines from $t_5$ to $t_4$ and also from $t_3$ to $t_2$.  At these poles the amplitude in the WKB solutions (\ref{WKB}) goes to zero which means that after going through these poles we have to start with an arbitrary linear combination of the WKB solutions (\ref{WKB}).  In other words we lose the information imposed by the boundary condition at infinity as soon as we cross the poles.  We are not aware of any other method which can make it possible to solve for the highly damped QNM frequency in the Schwarzschild spacetime with G-B correction.

\begin{figure}[tb]
\begin{center}
\includegraphics[height=10cm]{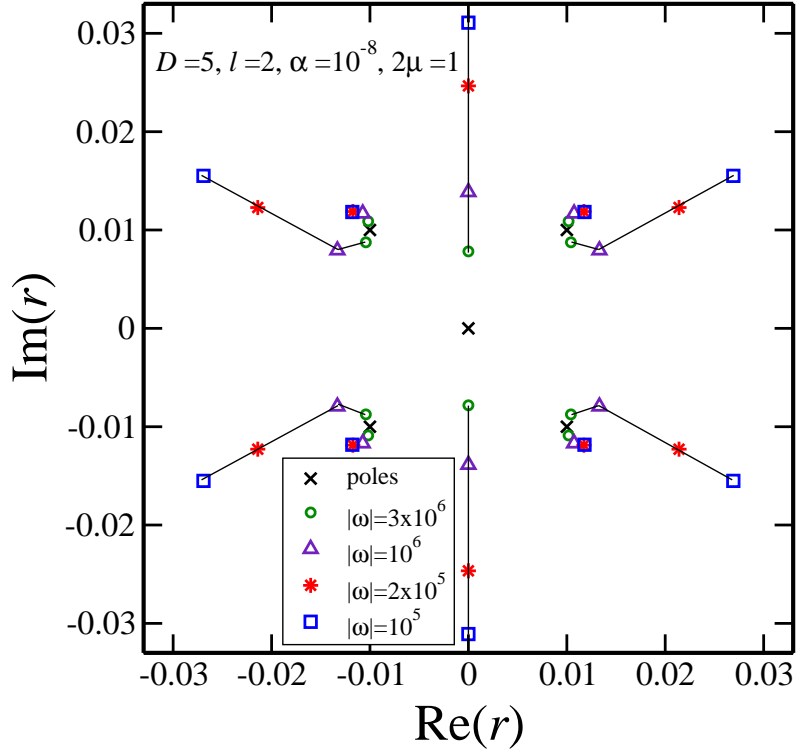}
\end{center}
\caption{The zeros of the function $Q$ in five spacetime dimensions for different values of $|\omega|$.}
\label{zeros-5d}
\end{figure}

\begin{figure}[tb]
\begin{center}
\includegraphics[height=11cm]{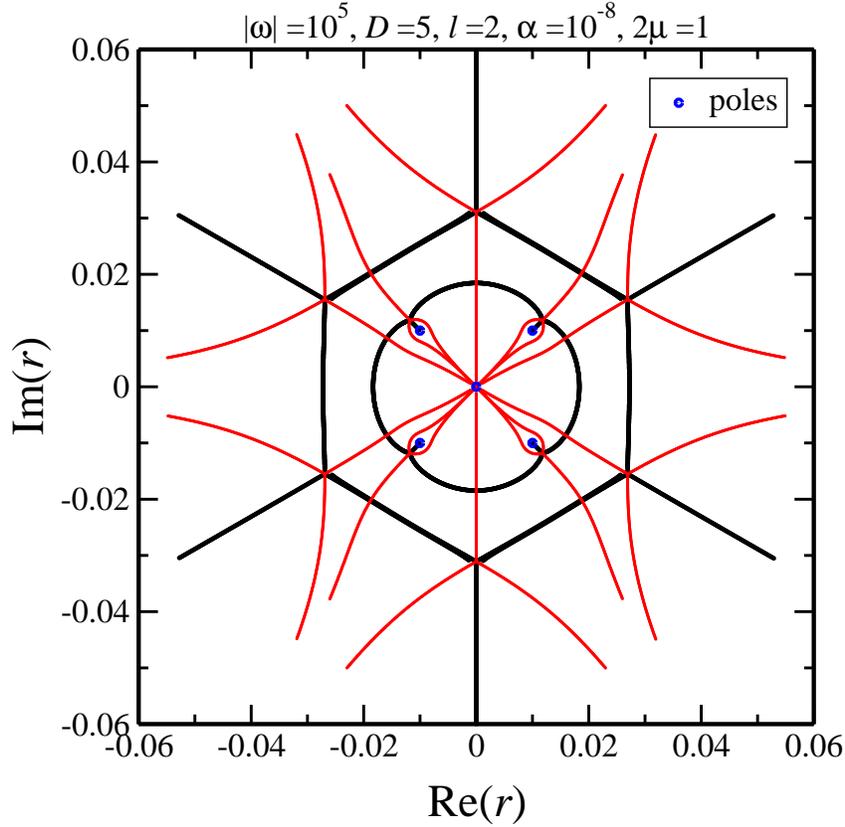}
\end{center}
\caption{Stokes (thin) and anti-Stokes (thick) lines produced numerically for vector perturbations in five spacetime dimensions.}
\label{numeric}
\end{figure}

\begin{figure}[tb]
\begin{center}
\includegraphics[height=10cm]{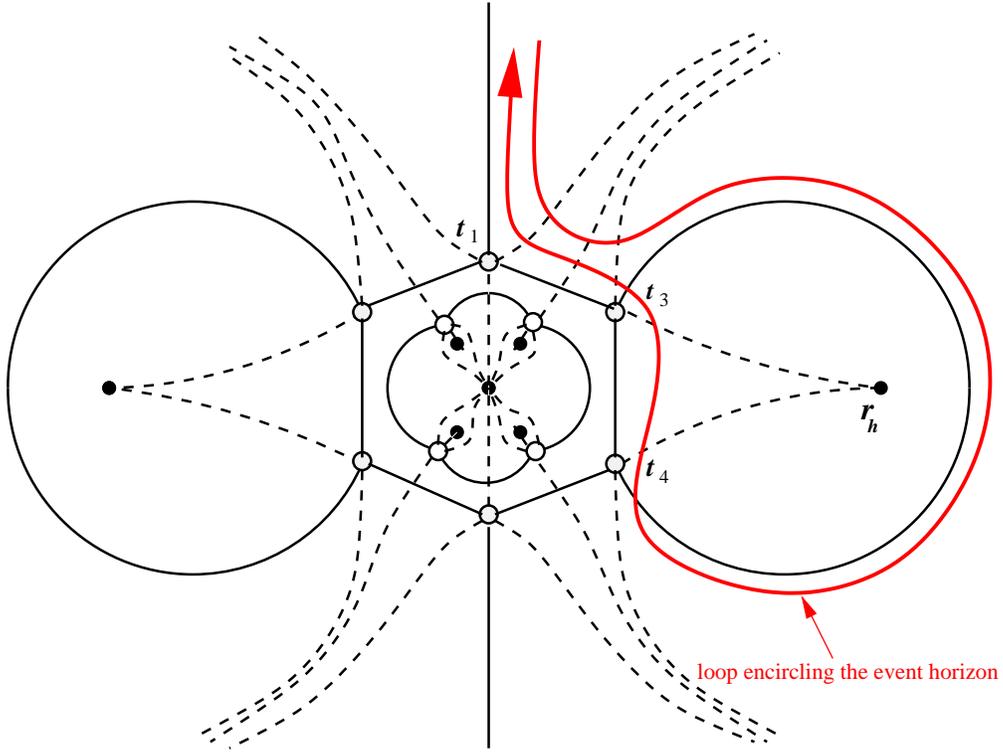}
\end{center}
\caption{A schematic illustration of the Stokes (dashed) and anti-Stokes (solid) lines in the complex $r$-plane for the G-B problem in five spacetime dimensions for an intermediate damping.  The open circles represent the zeros, while the filled circles are the poles of $Q^2$.}
\label{schemGB1}
\end{figure}

\begin{figure}[tb]
\begin{center}
\includegraphics[height=7cm]{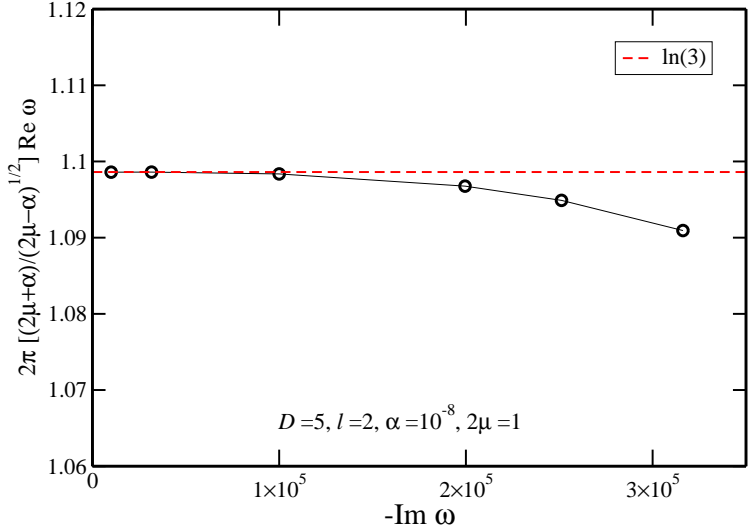}
\end{center}
\caption{$2 \pi {2\mu+\alpha \over \sqrt{2\mu-\alpha}}~{\mbox {Re}}~\omega$ as a function of $- {\mbox {Im}}~\omega$ for vector perturbations in five spacetime dimensions.}
\label{Re-Im-omega}
\end{figure}

\sxn{Vector perturbations in the intermediate damping region}

In this section, we investigate the approximation suggested in \cite{GB-asymp-qnm}, namely
\beeq
f(r) \approx 1-{2 \mu \over r^{D-3}}+{4 \alpha \mu^2 \over r^{2D-4}}~.
\label{f-approx}
\eneq
This approximation is only valid if all WKB calculations are done in a region of the complex plane where ${8 \alpha \mu \over r^{D-1}} \ll 1$ or equivalently $r \gg (8 \alpha \mu)^{1/(D-1)}$.  Thus, all the relevant Stokes/anti-Stokes lines and the zeros of the function $Q$ along the contour in the WKB analysis should be far from the pole at the origin and at $r^{D-1}+8\alpha \mu=0$ and $(D-3)r^{D-1}+4\alpha \mu(D-5)=0$.  This implies that the approximation (\ref{f-approx}), forbids taking the infinitely damped limit of the QNM frequency because in this limit the zeros of the function $Q$ approach the origin where the approximation (\ref{f-approx}) becomes invalid.  The authors of \cite{GB-asymp-qnm} ignored this crucial point in their calculations and consequently arrived at incorrect conclusions. 

However, if we lower the damping, the zeros of the function $Q$ will move away from the poles. 
 Fig. \ref{zeros-5d} shows how the zeros of the function $Q$ in five spacetime dimensions move in the complex plane as the damping decreases.  In this figure $2\mu=1$, $\alpha=10^{-8}$, and $l=2$.  As one can see, six of the zeros of the function $Q$ will move significantly outward away from the poles in the complex plane.  Four of the zeros will stay relatively close to the poles at $r^{4}+8\alpha \mu=0$.  Thus, for intermediate damping, the six zeros move to a region where $r \gg (8 \alpha \mu)^{1/4}$ and the approximation (\ref{f-approx}) is valid. It is easy to show that, in general, for the approximation (\ref{f-approx}) to be valid we need to have
\beeq
|\omega^2| \approx [{\rm Im}~ \omega]^2 \ll {(4n^2-1)\mu^2 \over (8\alpha \mu)^{2n \over n+1}}~.
\label{omega-condition}
\eneq
Note that as in the purely Schwarzschild calculation, we also require that the damping to be large compared to the real part of the frequency, which implies that 
\beeq
|\omega|>>{1\over \mu}
\eneq
and guarantees that the zeros along the contour used are well inside the horizon radius.

In this intermediate damping region the topology of the Stokes/anti-Stokes lines will look very different from the one in Fig. \ref{schematicGB}.  Fig. \ref{numeric} shows the numerically generated topology of the Stokes/anti-Stokes lines for five dimensional G-B black holes for an intermediate damping of $|\omega|=10^5$.  In this figure $2\mu=1$, $\alpha=10^{-8}$, and $l=2$.  To generate this figure we have written a numerical recipe that uses the approximation
\beeq
\int_{a}^{a+dz} \sqrt{\tilde{Q}(r)}dr = {2\over 3}{1\over \tilde{Q}'(a)}\left[(\tilde{Q}(a)+\tilde{Q}'(a)dz)^{3/2}-\tilde{Q}(a)^{3/2}\right]~,
\label{numeric1}
\eneq
when the stepsize $|dz|$ is small enough.  Here $\tilde{Q}=Q^2$.  The numerical calculation starts at the zeros of the function $Q$. To find the Stokes lines, the program finds the point where
\beeq
-\epsilon \leq {\rm Re} \left\{ {2\over 3}{1\over \tilde{Q}'(a)}\left[(\tilde{Q}(a)+\tilde{Q}'(a)dz)^{3/2}-\tilde{Q}(a)^{3/2}\right] \right\} \leq \epsilon~,
\label{numeric2}
\eneq 
and to find the anti-Stokes lines the program finds the points where
\beeq
-\epsilon \leq {\rm Im} \left\{ {2\over 3}{1\over \tilde{Q}'(a)}\left[(\tilde{Q}(a)+\tilde{Q}'(a)dz)^{3/2}-\tilde{Q}(a)^{3/2}\right] \right\} \leq \epsilon~.
\label{numeric3}
\eneq 
Here $\epsilon$ ideally needs to be zero, but in the numerical calculations we take it to be a small number which generally works well if we take it to be roughly equal to the stepsize $|dz|$.

We plot the schematic behavior of the Stokes/anti-Stokes lines in Fig. \ref{schemGB1} for G-B black holes in five spacetime dimensions in the intermediate damping region.  Note that the topology of the Stokes/anti-Stokes lines in this damping region looks almost identical to the topology of these lines for five dimensional Schwarzschild black holes.  The only region of the complex plane which is different from the Schwarzschild problem is the region contained inside the anti-Stokes lines connecting the outer six zeros.  This region stays outside of the loop, shown in Fig. \ref{schemGB1}, that we follow to extract the WKB condition and therefore will not affect the QNM calculations.  In addition, compared to the Schwarzschild problem,  there exist four extra Stokes lines in Fig. \ref{schemGB1} that extend to infinity.  These extra Stokes lines, which cross the anti-Stokes lines, will not affect our calculations either, because the exponential dominance of the WKB solutions (\ref{WKB}) only changes at the so called ``transition points'' which are the zeros of the function $Q$ for this particular problem.  More specifically, along the anti-Stokes line connecting $t_1$ to $t_3$ there will be no change in the dominance of the WKB solutions due to the crossing of the Stokes line.

Let us see how the function $R$ is modified if we use the approximate metric function $f(r)$ in Eq. (\ref{f-approx}).  In the large but finite damping limit and when $\alpha/\mu \ll 1$, the function $R$ takes the form
\beeq
R(r)={1\over f^2}\left \{ \omega^2-{(4n^2-1)\mu^2 \over r^{2n}}\left[ 1+{n^3-9n^2+18n+10 \over 2(4n^2-1)(n-1)}\left( {8\alpha \mu \over r^{n+1}}\right)+O\left[ \left( {8\alpha \mu \over r^{n+1}}\right)^2 \right] \right] \right\}~,
\label{Q-approx}
\eneq 
where $n=D-2$.  It is easy to see that as long as the calculations are done in a region of the complex plane where $r \gg (8\alpha \mu)^{1/n+1}$, we get the same function $R$ as in Schwarzschild black holes plus the leading terms which add a small correction due to the presence of the G-B coupling constant $\alpha$.

In Fig. \ref{Re-Im-omega}, we plot the QNM frequency as a function of damping.  Note that according to this figure the deviation of the real part of the frequency from $\ln(3)$ due to G-B corrections is very small.  To generate the numerical data in this figure, we have followed the same combination of analytic and numeric techniques developed in \cite{DKOB}.  First, we use the analytic method of {\cite{Andersson}} to find the general WKB condition
\beeq
e ^{2i\Gamma}=-e^{2i\gamma _{43}}-{(1+e ^{2i\gamma_{43}}) \over e ^{2i\gamma_{13}}}~
\label{WKB-Sch-generic}
\eneq 
for the Schwarzschild-type topology shown in Fig. \ref{schemGB1}.  Here
\beeq
\gamma_{ab}=\int^{t_b}_{t_a} Q(r)dr~
\label{gab}
\eneq 
and $\Gamma$ is the integral of the function $Q$ along a contour encircling the pole at the event horizon in the negative direction.  Using the residue theorem we can evaluate $\Gamma$ in five spacetime dimensions:
\beeq
\Gamma=\oint Q dy= -2\pi i \mathop{Res}_{r=|r_h|} Q= -2\pi i\left[{\frac{2\mu+\alpha}{2\sqrt{2\mu-\alpha}}\omega}\right] ~.
\label{Gamma}
\eneq
The phase integrals $\gamma_{13}$ and $\gamma_{43}$ are calculated numerically using the approximation (\ref{numeric1}).  For the particular case in Fig. \ref{numeric} we find $\gamma_{13}\sim -3.12853\cdots$ radians and $\gamma_{43}\sim -3.16878\cdots$ radians.  Therefore we get
\beeq
e ^{2i\Gamma}=e^{2\pi {\frac{2\mu+\alpha}{\sqrt{2\mu-\alpha}}\omega}}\approx -2.99494\cdots+i0.16088\cdots~.
\label{WKB-numeric1}
\eneq 
In other words 
\beeq
2\pi \frac{2\mu+\alpha}{\sqrt{2\mu-\alpha}}~\omega  \approx \ln(2.99926\cdots)-i(2n+1)\pi~,
\eneq
where $n \gg 1$.  In Fig. \ref{numeric}, we have taken $|\omega|\sim -{\mbox {Im}}~\omega =10^5$ which means that $n \approx 10^5$.

\sxn{Conclusion}

We have calculated, using a combination of analytic and numeric techniques, the QNM frequencies for
G-B black holes in the intermediate damping region, in which the
imaginary part of the QNM frequency is large relative to the black hole
mass, but not infinitely large.  Physically this corresponds to taking 
the low energy limit, $\alpha\to 0$  before $\omega\to \infty$. As
expected in this limit, the frequencies are perturbations of the
$\ln(3)$ that appears for the highly damped QNM perturbations of the
Schwarzschild black hole. Interestingly, the G-B corrections decrease the
real part of the frequency as the damping is increased.  Therefore, contrary to what was claimed by the authors of \cite{GB-asymp-qnm}, we conclude that the ``universal'' value of $\ln(3)$ with a small deviation due to G-B corrections does appear in G-B black holes.

The story will change for the infinite damping limit.  The infinitely
damped QNMs essentially ``see'' the full structure of the solution
including the extra poles of the function $Q$ which only exist in G-B
black holes.  These extra poles will affect the value of the frequency
in this limit.    This phenomenon is very similar to what happens in
Reissner-Nordstr$\ddot{\rm{o}}$m (R-N) black holes with small charge,
where the real part of the QNM frequencies makes a transition from R-N
value in the infinitely damped limit to the Schwarzschild value of
$\ln(3)T_{bh}$ for an intermediate damping \cite{DKOB}.  In that case
also the infinitely damped QNMs essentially ``see'' the full structure
of the solution, including the inner horizon, which affects the value of
the frequency in this limit. The intermediate damping on the other hand
only probes the outer horizon in R-N black hole, and for this range the
``universal'' value of $\ln(3)$ is reproduced. In the present context, this behavior suggests that in order to solve for the highly damped QNMs we may need to work with a more complete model of black holes in string theories rather than the G-B black hole which only includes the leading order higher curvature terms arising in the low energy limit of string theories. 

\vskip .5cm

\leftline{\bf Acknowledgments}
This research was supported in part by the Natural Sciences and
Engineering Research Council of Canada. We are grateful to Esteban
Herrera for useful conversations.


\def\jnl#1#2#3#4{{#1}{\bf #2} (#4) #3}

\def\Zphys{{\em Z.\ Phys.} }
\def\jssc{{\em J.\ Solid State Chem.\ }}
\def\jpsJ{{\em J.\ Phys.\ Soc.\ Japan }}
\def\ptps{{\em Prog.\ Theoret.\ Phys.\ Suppl.\ }}
\def\PTP{{\em Prog.\ Theoret.\ Phys.\  }}
\def\LNC{{\em Lett.\ Nuovo.\ Cim.\  }}

\def\JMP{{\em J. Math.\ Phys.} }
\def\NPB{{\em Nucl.\ Phys.} B}
\def\NP{{\em Nucl.\ Phys.} }
\def\PLB{{\em Phys.\ Lett.} B}
\def\PL{{\em Phys.\ Lett.} }
\def\PRL{\em Phys.\ Rev.\ Lett. }
\def\PRB{{\em Phys.\ Rev.} B}
\def\PRD{{\em Phys.\ Rev.} D}
\def\PR{{\em Phys.\ Rev.} }
\def\PRe{{\em Phys.\ Rep.} }
\def\AP{{\em Ann.\ Phys.\ (N.Y.)} }
\def\RMP{{\em Rev.\ Mod.\ Phys.} }
\def\ZPC{{\em Z.\ Phys.} C}
\def\SCI{\em Science}
\def\CMP{\em Comm.\ Math.\ Phys. }
\def\MPLA{{\em Mod.\ Phys.\ Lett.} A}
\def\IJMPB{{\em Int.\ J.\ Mod.\ Phys.} B}
\def\cmp{{\em Com.\ Math.\ Phys.}}
\def\JPA{{\em J.\  Phys.} A}
\def\CQG{\em Class.\ Quant.\ Grav.~}
\def\ATMP{\em Adv.\ Theoret.\ Math.\ Phys.~}
\def\PRSA{{\em Proc.\ Roy.\ Soc.} A }
\def\ibid{{\em ibid.} }
\vskip 1cm

\leftline{\bf References}

\renewenvironment{thebibliography}[1]
        {\begin{list}{[$\,$\arabic{enumi}$\,$]}  
        {\usecounter{enumi}\setlength{\parsep}{0pt}
         \setlength{\itemsep}{0pt}  \renewcommand{\baselinestretch}{1.2}
         \settowidth
        {\labelwidth}{#1 ~ ~}\sloppy}}{\end{list}}



\begin{thebibliography}{99}
\small
\baselineskip=16pt




\bibitem{Hod}
S. Hod, \jnl{\PRL}{81}{4293}{1998}.

\bibitem{Dreyer} O.~Dreyer, \jnl{\PRL}{90}{081301}{2003}.



\bibitem{Natario} 
J.~Natario and R.~Schiappa, \jnl{\ATMP}{8}{1001}{2004}; E. Berti, ''Black hole quasinormal modes: hints of quantum gravity?'', gr-qc/0411025; V.~Cardoso, J.~Natario and R.~Schiappa, \jnl{\JMP}{45}{4698}{2004};
 
 
\bibitem{Motl2}
L. Motl, \jnl{\ATMP}{6}{1135}{2003}; L. Motl and A. Neitzke, \jnl{\ATMP}{7}{307}{2003}.

\bibitem{Gabor3}
J. Kettner, G. Kunstatter, A.J.M. Medved, \jnl{\CQG}{21}{5317}{2004}.

\bibitem{Das1}
S. Das and S. Shankaranarayanan,
\jnl{\CQG}{22}{L7}{2005}.


\bibitem{Ramin1} R. Daghigh and G. Kunstatter, \jnl{\CQG}{22}{4113}{2005}.


\bibitem{GB1}D.~G.~Boulware and S. Deser, \jnl{\PRL}{55}{2656}{1985}.
\bibitem{GB2}J.~Wheeler, \jnl{\NPB}{268}{737}{1986}.
\bibitem{GB3}D.~L.~Wiltshire, \jnl{\PRD}{38}{2445}{1988}.

\bibitem{Zwiebach}B.~Zwiebach, \jnl{\PLB}{156}{315}{1985}.


\bibitem{GB-brane}A.~Barrau, J. Grain, and S.O. Alexeyev, \jnl{\PLB}{584}{114}{2004}.


\bibitem{GB-qnm1}B.~R.~Iyer, S. Iyer, and C. V. Vishveshwara, \jnl{\CQG}{6}{1627}{1989}.
\bibitem{GB-qnm2}R.~Konoplya, \jnl{\PRD}{71}{024038}{2005}.
\bibitem{GB-qnm3}E.~Abdalla, R. A. Konoplya, and C. Molina, \jnl{\PRD}{72}{084006}{2005}.
\bibitem{GB-qnm4}S. K. Chakrabarti, hep-th/0603123.


\bibitem{GB-asymp-qnm}S. K. Chakrabarti and K. S. Gupta, \jnl{JMPA}{21}{3565}{2006}. 

\bibitem{GB-asymp-qnm1}F. Moura and R. Schiappa, hep-th/0605001.

\bibitem{Deser}
D. G. Boulware and S. Deser, \jnl{\PRL}{55}{2656}{1985}.

\bibitem{Dotti1}
G. Dotti, R. J. Gleiserand, \jnl{\CQG}{22}{L1}{2005}.

\bibitem{Dotti2}
G. Dotti, R. J. Gleiserand, \jnl{\PRD}{72}{044018}{2005}.


\bibitem{Dotti3}
G. Dotti, R. J. Gleiserand, \jnl{\PRD}{72}{124002}{2005}.




\bibitem{Andersson}
N. Andersson and C. J. Howls, \jnl{\CQG}{21}{1623}{2004}.


\bibitem{DKOB}R. Daghigh, G. Kunstatter, D. Ostapchuk, and V. Bagnulo,\jnl{\CQG}{23}{5101}{2006}.









\end{thebibliography}
\end{document}